\documentclass[aps,prb,floatfix]{revtex4}

\usepackage{graphicx}
\usepackage{color}
\usepackage{amsbsy,amsmath}
\usepackage{bm}% bold math
\usepackage{array,multirow,makecell}
\usepackage{amsfonts}

\newcommand{\be}{\begin{equation}}
\newcommand{\ee}{\end{equation}}

\begin{document}
\title{High-harmonic generation in a quantum electron gas trapped in a nonparabolic and anisotropic well}

\author{J\'er\^ome Hurst, K\'evin L\'ev\^eque-Simon, Paul-Antoine Hervieux, Giovanni Manfredi}
\email{manfredi@unistra.fr}
\affiliation{Institut de Physique et Chimie des Mat\'eriaux de
Strasbourg, CNRS and Universit\'e de Strasbourg, BP 43, F-67034 Strasbourg Cedex 2, France }
\author{Fernando Haas}
\email{fernando.haas@ufrgs.br}
\affiliation{Physics Institute, Federal University of Rio Grande do Sul,  Avenida Bento Gon\c{c}alves 9500, CEP 91501-970, Porto Alegre, RS, Brazil}

\date{\today}

\begin{abstract}
An effective self-consistent model is derived and used to study the dynamics of an electron gas confined in a nonparabolic and anisotropic quantum well. This approach is based on the equations of quantum hydrodynamics, which incorporate quantum and nonlinear effects in an approximate fashion. The effective model consists of a set of six coupled differential equations (dynamical system) for the electric dipole and the size of the electron gas. Using this model we show that: (i) High harmonic generation is related to the appearance of chaos in the phase space, as attested by related Poincar\'e sections; (ii) Higher order harmonics can be excited efficiently and with relatively weak driving fields by making use of chirped electromagnetic  waves.
\end{abstract}

\maketitle

\section{Introduction}\label{sec:intro}

Current technology allows the manipulation and control of the electron dynamics in small devices of nanometric size, such as semiconductor quantum dots and quantum wells. These devices have attracted considerable attention in the last few decades, particularly in view of their potential use for quantum computing \cite{Zoller}.

When the confining potential well is perfectly parabolic, the electron
response is dominated by the Kohn mode \cite{Kohn, Dobson}, i.e.,
a rigid oscillation of the electron density at the characteristic frequency of the parabolic well.
For nonparabolic confinement the situation is much more complex. When the excitation is small (linear response), the Kohn mode may still be dominating. However, for larger excitation energies, the electrons may explore the
anharmonic regions of the confining potential; in that case, the frequency spectrum of such nonlinear response becomes much more intricate, with the appearance of second- and higher-order harmonics.
In addition to the effect of the anharmonicity of the confinement, the interparticle Coulomb interactions also contribute to the complexity of the spectral response.
At a mathematical level, this complexity arises because the center-of-mass and internal degrees of freedom can no longer be separated, as was shown in several studies that use powerful exact methods to model the quantum electron dynamics \cite{sako,schroter}.

However, exact approaches are necessarily limited to a very small number of particles. Although such few- or even single-electron systems can nowadays be realized in the laboratory, in most practical situations a great many electrons are involved \cite{Muller,Pereira}.
In that case self-consistent effects -- arising from the Coulomb interactions between all the electrons -- play a crucial role on the dynamics.
Several theoretical and computational studies, which treat the many-body dynamics in an approximate way, have investigated
the linear and nonlinear electron response. The methods of choice are the Hartree-Fock equations \cite{Nikonov}, density functional theory (DFT)
\cite{Ullrich}, or phase-space methods based on Wigner functions \cite{ap_auto4}.

Even the above-cited methods can be computationally too costly for very large systems. A possible alternative relies on quantum hydrodynamics (QHD) \cite{fernando_book,qhy}, an approach that was successfully used in the past to model the electron dynamics in molecular systems \cite{qhy_mol}, metallic nanoparticles \cite{qhy_clust1,qhy_clust2,manfredi2012,ciraci} and thin films \cite{qhy_thin_met}, and semiconductor quantum wells \cite{qhy_semi_cond}.
In Ref. \cite{qhy_thin_met} the validity of the QHD method was studied and compared to DFT results.
More recently, we used a QHD approach to investigate  high-harmonic generation in metallic nanoparticles excited with ultrafast laser pulses \cite{hurst_harmo}.

The QHD model can be further simplified by means of a variational approach \cite{manfredi2012} that expresses the QHD equations in terms of a Lagrangian density. With this method, it is possible to obtain a  system of ordinary differential equations for a set of macroscopic quantities, such as the center of mass and the size of the electron gas. Although simple, the final equations still capture some of the most prominent features of the electron dynamics, namely: (i) the self-consistent Coulomb interaction, (ii) quantum effects to lowest order, (iii) exchange and correlation effects in a DFT fashion, and (iv) the geometry of the confining well.

Here, we will use this approach to study the collective  dynamics of an electron gas confined in a semiconductor quantum well. The simplicity of the model allows one to carry out a large number of simulations, so that the electron dynamics can be fully characterized.
Our main focus will be on the effect of the anharmonicity and the anisotropy of the potential well on the electron response. We will see that, when increasing the anharmonic component of the confining potential, the electron dynamics becomes more and more complex and eventually fully chaotic. The anisotropy of the confinement and the magnitude of the Coulomb effects (i.e., the number of trapped electrons) also play an important role in this transition. We will finally show that the appearance of chaotic behavior is accompanied by the presence of higher-order harmonics in the electron response.

\section{QHD Model}\label{sec:model}
The set of QHD equations for the electron density $n(\bm{r},t)$, the electron mean velocity $\bm{u}(\bm{r},t)$, and the Hartree potential $V_H(\bm{r},t)$ reads as
\begin{align}
&\frac{\partial n}{\partial t} + \nabla\!\cdot(n\bm{u})=0, \label{eq:continuity}\\
&\frac{\partial \bm{u}}{\partial t}+\bm{u}\cdot\!\nabla \bm{u}=\nabla\!\,V_{H}-\nabla\!\,V_{\rm conf}-\nabla\!\,V_{X}-\frac{\nabla P}{n}+\frac{1}{2}\nabla \left(\frac{\nabla^2\sqrt{n}}{\sqrt{n}}\right), \label{eq:momentum}\\
&\Delta\,\!V_H=4\pi n, \label{eq:poisson}
\end{align}
where the first equation above represents conservation of mass, the second represents conservation of momentum, and the third is Poisson's equation for the self-consistent Hartree potential $V_H$.
In Eq. \eqref{eq:momentum}, $V_{\rm conf}(\bm{r},t)$ is the potential of the confining well, $V_X$ is the exchange potential (see below), and $P(\bm{r},t)$ is the Fermi pressure of a degenerated electron gas (we will make the assumption that the system's temperature is always much lower than the Fermi temperature)
\begin{equation}
P=\frac{1}{5}\left(3\pi^2\right)^{2/3}n^{5/3}.
\end{equation}
The last term in Eq. \eqref{eq:momentum} is the so-called Bohm potential, which incorporates quantum effects to the lowest order. The Bohm potential is related to the so-called von Weizs\"acker term in Thomas-Fermi theory and orbital-free DFT \cite{Bonitz2015}.

The above equations are written in ``semiconductor" atomic units (au). These are formally identical to standard au, but the electron mass $m_e$ is replaced by the effective mass $m^{\displaystyle*}=0.067m_{e}$ and the vacuum dielectric constant $\epsilon_0$ by its effective counterpart $\epsilon^{\displaystyle*}=13 \epsilon_{0}$.
In this system of units, length are normalized to an effective Bohr radius $a^{\displaystyle*} = 4\pi \epsilon^{\displaystyle*}\hbar^{2}/(m^{*}e^{2})= 10.3\, \rm nm$, energy to an effective Hartree energy  $E_{H}^{\displaystyle*} = \hbar^{2}/(m^{\displaystyle *}a^{{\displaystyle*}2}) = 10.8\, \rm meV$, frequency to $\omega^{\displaystyle*} = E_{H}^{\displaystyle*}/\hbar =2\pi \times 2.63\, \rm THz$, and time to $\tau^{\displaystyle*}= 1/\omega^{\displaystyle*} = 0.061 \, \rm ps$.
These units will be used throughout this work unless otherwise specified.

The confining potential is given by the sum of a harmonic and an anharmonic (but isotropic) part, whose relative strength is measured by the parameter $\zeta \ge 0$:
\begin{equation}
V_{\text{conf}}=\frac{1}{2}\left(k_xx^2+k_yy^2+k_zz^2\right) + \zeta \left(x^2+y^2+z^2\right)^{2} \label{eq:Vconf}.
\end{equation}
We chose this specific form for the anharmonic part of the confinement, so that it can be captured by the single parameter $\zeta$.
The elastic constants $k_{i}$ of the harmonic potential in Eq. \eqref{eq:Vconf} are normalized to $k^{\displaystyle*} \equiv \omega^{{\displaystyle*}\,2}m^{\displaystyle *}= 1.64 \times 10^{-5}\rm J/m^{2}$.

As in DFT, exchange effects can be modeled by a density-dependent effective potential
\begin{equation}
V_{X}[n(\bm{r},t)]=-\frac{1}{\pi}\left(3\pi^2 n\right)^{1/3}+\beta\Bigg[2\,\frac{\nabla^2\,\!n}{n^{4/3}}-
\frac{4}{3}\frac{\big(\nabla\,\!n\big)^2}{n^{7/3}}\Bigg],
\end{equation}
where the first term is the local density approximation (LDA) and the other two terms constitute a gradient correction.
The prefactor $\beta$ is a free parameter that we set equal to $\beta = 0.005$, which is a best-fit frequently used in atomic-structure calculations \cite{corr_pot,manfredi2012}.

The QHD equations \eqref{eq:continuity}-\eqref{eq:poisson} can be represented, without further approximations, by a Lagrangian density $\mathcal{L}(n,\theta,V_H)$, where the function $\theta(\bm{r},t)$ is related to the mean velocity, $\bm{u} = \nabla \theta$. The expression for the Lagrangian density is as follows:
\begin{align}
\mathcal{L}=\,&n\bigg[\frac{1}{2}\big(\nabla\,\!\theta\big)^2\!+\frac{\partial \theta}{\partial t}\bigg]+\frac{1}{8n}\big(\nabla\,\!n\big)^2+\
\frac{3}{10}\big(3\pi^2\big)^{2/3}n^{5/3}\nonumber\\[-2.5mm]\label{eq.35}\\[-2.5mm]
&-\frac{3}{4\pi}\big(3\pi^2\big)^{1/3}n^{4/3}-
\beta\frac{\big(\nabla\,\!n\big)^2}{n^{4/3}}+
nV_{\text{conf}}-nV_{H}-\frac{1}{8\pi}\big(\nabla\,\!V_H\big)^2\nonumber .
\label{eq:lagrange-dens}
\end{align}
By taking the standard Euler-Lagrange equations with respect to the three fields
$n$, $\theta$, and $V_H$, one recovers exactly the QHD equations \eqref{eq:continuity}-\eqref{eq:poisson}.

In order to derive a tractable system of equations, one needs to specify a particular Ansatz for the electron density. Here, we take a Gaussian shape, which is a reasonable choice, as it is the exact ground-state solution when one neglects Coulomb interactions and the anharmonic part of the confinement.
Thus we write:
\begin{equation}
n(\bm{r},t)=\frac{A}{\sigma_x\sigma_y\sigma_z}\exp\Big(-\frac{1}{2}\rho^2\Big), \label{eq:density}
\end{equation}
where the prefactor $A = N/\left(2\pi\right)^{3/2}$ is obtained by fixing the total number of particles $N=\int n d\bm{r}$, and $\rho$  is a displaced position variable
\begin{equation}
\rho(x,y,z,t)=\sqrt{\frac{1}{\sigma_x^2}\big(x-d_x\big)^2+\frac{1}{\sigma_y^2}\big(y-d_y\big)^2+
\frac{1}{\sigma_z^2}\big(z-d_z\big)^2}\label{eq:rho}.
\end{equation}
Here, $d_i(t)$ and $\sigma_{i}(t)$ are time-dependant variables that represent respectively the center of mass and the size of the electron gas in each Cartesian direction.

We now need to express the other variables ($\theta$ and $V_H$) in terms of the electron density. The mean velocity $\bm{u}$ can be obtained exactly from the continuity equation \eqref{eq:continuity}. Its Cartesian components are:
\begin{equation}
u_i=\frac{\dot{\sigma_i}}{\sigma_i}\big(r_i-d_i\big)+
\dot{d_i}\label{eq:ui},
\end{equation}
where $i=\{x,y,z\}$ and $r_i=(x,y,z)$. From the above expression we obtain
\begin{equation}
\theta=\sum_{i}\left(\frac{\dot{\sigma_i}}{2\sigma_i}
(r_i-d_i)^2+\dot{d_i}(r_i-d_i)\right)
\label{eq:theta}.
\end{equation}

For the self-consistent Hartree potential, we take the expression
\begin{equation}
V_H = -4\pi\sqrt{\frac{\pi}{2}}\frac{A}{\left(\sigma_x\sigma_y\sigma_z\right)^{1/3}}
\,\frac{\rm erf\left(\rho/\sqrt{2}\right)}{\rho} \label{eq:VH},
\end{equation}
where $\rm erf$ is the error function. This expression constitutes an approximate solution of Poisson's equation \eqref{eq:poisson}, which becomes exact in the radially symmetric case ($\sigma_x=\sigma_y=\sigma_z$).

Substituting the above expression for $n$, $\theta$, and $V_H$ into the Lagrangian density \eqref{eq.35} and integrating over the entire space, we obtain the following Lagrangian function:
\begin{equation}
L[d_i,\sigma_{i},\dot d_i, \dot \sigma_{i}]=\frac{1}{N}\int \mathcal{L} d\bm{r}=\frac{1}{2}\sum_{i}\left(\dot{\sigma_{i}}^{2}+
\dot{d_i}^{2}\right)- U(d_i,\sigma_i)\label{eq:lagrangian},
\end{equation}
where a dot stands for differentiation with respect to time
and $U(d_{i},\sigma_i)=U_d(d_i)+U_{\sigma}(\sigma_i)+U_{d \sigma}(d_{i},\sigma_i)$. The different potential terms read as:
\begin{align}
U_d=&\frac{1}{2}\sum_{i}k_i d_{i}^{2},\label{eq:Ud}\\
U_{\sigma}=&\nonumber\frac{1}{2}\sum_{i}k_i\sigma_{i}^{2} + \left(\sum_{i}\frac{1}{\sigma_{i}^{2}}\right)\left(\frac{1}{8}+
\alpha_1N\big[\sigma_x\sigma_y\sigma_z\big]^{1/3}\!-
\,\alpha_2\,\beta\bigg[\frac{\sigma_x\sigma_y\sigma_z}{N}\bigg]^{1/3}\right)\label{eq:Usig}\\
&+\,\alpha_3\bigg[\frac{N}{\sigma_x\sigma_y\sigma_z}\bigg]^{2/3}\!\!-
\,\alpha_4\bigg[\frac{N}{\sigma_x\sigma_y\sigma_z}\bigg]^{1/3},\\
U_{d\sigma}=&\zeta\Bigg[\sum_{i}\Big(3\hspace{0.3mm}\sigma_{i}^4+
6\hspace{0.3mm}d_{i}^2\sigma_{i}^2+d_{i}^4\Big)\,+
\sum_{i\ne k}\!\!\Big(\sigma_{i}^2+
d_{i}^2\Big)\Big(\sigma_{k}^2+d_{k}^2\Big)\Bigg]\label{eq:Udsig},
\end{align}
and represent respectively the dipole motion ($U_d$), the breathing motion ($U_{\sigma}$), and the coupling between the dipole and breathing dynamics ($U_{d\sigma}$). Note that such coupling disappears for purely harmonic confinement ($\zeta=0$).
The various coefficient appearing in Eqs. \eqref{eq:Ud}-\eqref{eq:Udsig} are given by:
\begin{align}
&\alpha_1=\frac{2\pi}{3\sqrt{2}}\big(2\pi\big)^{-3/2} \approx 0.0940,\nonumber\\
&\alpha_2=\frac{9}{4}\sqrt{3\pi}\approx 6.9075,\nonumber\\
&\alpha_3=\frac{9}{50}\sqrt{\frac{3}{5}}\,\big(3\pi^2\big)^{2/3}\frac{1}{2\pi} \approx 0.2124,\nonumber\\
&\alpha_4=\frac{9\sqrt{3}}{32\pi}\frac{\big(3\pi^2\big)^{1/3}}{\sqrt{2\pi}}\approx 0.1914.\nonumber
\end{align}
Finally, the equations of motion of the system can be obtained from the Euler-Lagrange equations for $L$, and read as:
\begin{align}
&\ddot{d_i}=-\frac{\partial U_{d}}{\partial d_i}-\frac{\partial U_{d\sigma}}{\partial d_i},~~~~~
\ddot{\sigma_i}=-\frac{\partial U_{\sigma}}{\partial \sigma_i}-\frac{\partial U_{d\sigma}}{\partial \sigma_i}. \label{eq:euler-lagrange}
\end{align}
As expected, in the case of harmonic confinement ($\zeta=0$) the dipole and breathing modes are completely decoupled (Kohn's theorem \cite{Kohn, Dobson}).

We have thus reduced the complex problem of the dynamics of a multi-electron system to a relatively simple system of six coupled differential equations for the center of mass  and size of the electron gas, which can be solved on a desktop computer using standard methods (e.g., Runge-Kutta). As noted in the introduction, this approximate system still incorporates such important effects as Coulomb interactions, quantum and exchange effects, as well as the effects of the geometry of the confining trap (anharmonicity and anisotropy). Also, no assumptions of linearity were made, so that Eqs. \eqref{eq:euler-lagrange} can be used to study the nonlinear response of the electron gas.

\section{Ground state and linear regime}
Stationary states are obtained by setting $\ddot{d_{i}}=\ddot{\sigma_{i}}=0$. For the dipole mode, the solutions are clearly $d_{i}=0$. For the breathing mode,
the equations of motions are those of a fictitious particle evolving in the external potential $U_{\sigma} + U_{d\sigma}$. The equilibrium solution $\sigma_i^{(0)}$ corresponds to the minimum of such potential and can be found by setting its first derivative to zero:
\begin{align}
\left. \frac{\partial \left(U_{\sigma} + U_{d\sigma}\right)}{\partial \sigma_{i}} \right|_{d_{i}=0} &= k_{i} \sigma_{i} - \frac{2}{\sigma_{i}^{3}}\left[\frac{1}{8} + \gamma_{2} V^{1/3}\right] + q \frac{\gamma_{2} }{3} \left[ \frac{V^{1/3}}{\sigma_{i}}\right] -  \frac{2\gamma_{1} }{3} \left[ \frac{V^{-2/3}}{\sigma_{i}}\right] +  \frac{\gamma_{3} }{3} \left[ \frac{V^{-1/3}}{\sigma_{i}}\right]\nonumber \\
&~~~+ \zeta \left[ 12 \sigma_{i}^{3} + 4 \sigma_{i} \left( \sum_{j\neq i} \sigma_{j}^{2} \right) \right] = 0,
\label{minimum_potential}
\end{align}
where $V(t) = \sigma_x \sigma_y \sigma_z$ is the electron gas volume, $q= \sum_{i} \sigma_{i}^{-2}$, and the following additional parameters were defined $\gamma_{1} = \alpha_{3}N^{2/3}$, $\gamma_{2}=\alpha_{1}N - \alpha_{2}\beta N^{-1/3}$ and $\gamma_{3} = \alpha_{4}N^{1/3}$.

\begin{figure}[h!]
        \centering \includegraphics[scale=0.4]{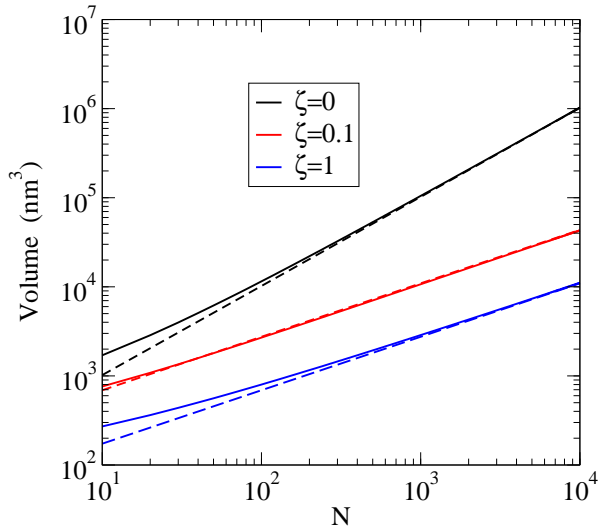}
        \caption{\textit{Color online}. Volume $V = \sigma_x \sigma_y \sigma_z = \sigma^{(0)\,3}$ of the electron gas as a function of the number of electrons $N$ for an isotropic confinement and different values of the anharmonicity parameter $\zeta$. The solid lines are solutions of Eq. (\ref{minimum_potential}), whereas the dashed lines are analytical solutions obtained in the $N \to \infty$ limit.}
        \label{fig:volume}
\end{figure}

In the most general case, it is not possible to obtain analytical solutions for the ground state and one has to resort to numerical methods. However, an analytical solution can be found in the case of isotropic confinement ($k_{x}=k_{y}=k_{z}\equiv k$) and in the limit of a large number of particles ($N \gg 1$).
We found that the size of the electron gas scales as a power of the number of electrons $N$. The exponent varies according to whether the anharmonic part of the potential is included or not: $\sigma^{(0)}=\left(\frac{\alpha_{1}}{k}N\right)^{1/3}$ for $\zeta=0$ (harmonic) and  $\sigma^{(0)}=\left(\frac{\alpha_1}{20 \zeta}N\right)^{1/5}$ for $\zeta\neq 0$ (anharmonic). As expected, for harmonic confinement the volume increases as the total number of particles. For the anharmonic case, the increase with $N$ is slower, reflecting the fact that the anharmonic term tends to further confine the electrons. Note that in this case the exponent is always $1/5$, irrespective of the value of $\zeta$. These results were confirmed by numerical simulations of the isotropic case obtained without making the large-$N$ approximation (Fig. \ref{fig:volume}).

Having found the ground state, it is possible to compute the linear response frequencies of the system. In the most general case, there are six such frequencies,
three of which correspond to the dipole (center-of-mass) modes $\Omega_{d}$ and three corresponding to the breathing modes $\Omega_{\sigma}$. These frequencies are obtained by finding the eigenvalues of the Hessian matrix constructed out of the second derivatives of the potential
\begin{align}
\mathcal{H} =
\begin{pmatrix}
\mathcal{H}_{\sigma} &0\\
0&  \mathcal{H}_{d}
\end{pmatrix},
\end{align}
where
\begin{align}
&\mathcal{H}_{\sigma} = {\frac{\partial ^{2} U}{\partial \sigma_{i} \partial\sigma_{j}} \vline}_{~d_i=0, ~\sigma_i=\sigma_i^{(0)}}, \nonumber\\
&  \mathcal{H}_{d} = {\frac{\partial^{2} U}{\partial d_{i} \partial d_{j}}\vline}_{~d_i=0, ~\sigma_i=\sigma_i^{(0)}} \nonumber
\end{align}
are symmetric ($\mathcal{H}_{\sigma}$) and diagonal ($\mathcal{H}_{d}$) $3\times 3$ matrices.

\begin{figure}[h!]
        \centering \includegraphics[scale=0.4]{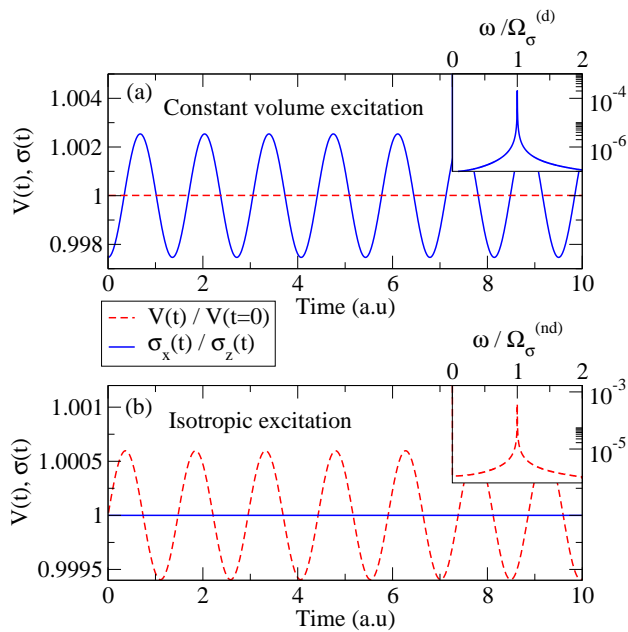}
        \caption{\textit{Color online}. Simulations of the two breathing modes in the linear regime for an isotropic case $k_{x}=k_{y}=k_{z}=1$ with $N=50$ and $\zeta=0.1$. Top panel (a): degenerate mode with constant volume $V(t)$ (red dashed curve) and varying aspect ratio $\sigma_{x}/\sigma_{z}$ (blue continuous curve); Bottom panel (b): non-degenerate mode with constant aspect ratio and oscillating volume. The insets show the Fourier transforms of the oscillating quantities, which peak at the expected degenerate  ($\Omega_{\sigma}^{(d)}$) and non-degenerate ($\Omega_{\sigma}^{(nd)}$) frequencies.}
        \label{fig:linearmodes}
\end{figure}

\begin{table}
\begin{tabular}{|c|c|c|c||c|c|c|c|c|}
  \hline
   & & & &  &\\[-3mm]
$N$ & $k_x$ & $k_y$ & $k_z$  & $\hspace{1.1cm}\Omega_{\sigma}/\omega^{*}$ & $\hspace{1.1cm}\Omega_{d}/\omega^{*}$
  \\[1mm] \hline
  \multirow{4}{*}{20}
    & & & & &\\[-2mm]
    &1&3&2&3.21\hspace{3mm}3.69\hspace{3mm}4.19&1.43\hspace{3mm}1.73\hspace{3mm}1.99\\[1mm]
    &1&1&2&3.17\hspace{3mm}3.32\hspace{3mm}3.76&1.46\,\,\,\,($\times$2)\hspace{2.4mm}1.75\\
    &1&1&1&3.14\hspace{3mm}3.34\,\,\,\,($\times$2)&1.46\,\,\,\,($\times$3)\\[1mm] \hline
  \multirow{4}{*}{50}
    & &  & & &\\[-2mm]
    &1&3&2&3.63\hspace{3mm}4.15\hspace{3mm}4.64&1.60\hspace{3mm}2.10\hspace{3mm}1.87\\[1mm]
    &1&1&2&3.55\hspace{3mm}3.88\hspace{3mm}4.25&1.63\,\,\,\,($\times$2)\hspace{2.4mm}1.89\\[1mm]
    &1&1&1&3.51\hspace{3mm}3.90\,\,\,\,($\times$2)&1.65\,\,\,\,($\times$3)\\[1mm]\hline
  \multirow{4}{*}{100}
    & &  & & &\\[-2mm]
    &1&3&2&4.00\hspace{3mm}4.61\hspace{3mm}5.07&1.76\hspace{3mm}2.00\hspace{3mm}2.23\\[1mm]
    &1&1&2&3.92\hspace{3mm}4.38\hspace{3mm}4.70&1.79\,\,\,\,($\times$2)\hspace{2.4mm}2.03\\[1mm]
    &1&1&1&3.89\hspace{3mm}4.40\,\,\,\,($\times$2)&1.81\,\,\,\,($\times$3)\\[1mm] \hline
\end{tabular}
\caption{Dipole ($\Omega_d$) and breathing ($\Omega_{\sigma}$) frequencies for various numbers of particles $N$ and different geometries of the confinement. The anharmonicity constant is $\zeta=0.05$ everywhere.}
\end{table}

In the following, we concentrate on the effect of the anharmonicity and focus on the case $\zeta=0.05$.
The numerically computed linear frequencies are given in Tab. I for different confinements (isotropic and anisotropic) and different numbers of particles.
For an isotropic case ($k_x=k_y=k_z=1$) we obtain, as expected, a single value for the dipole frequency that is three times degenerate, but two values for the breathing frequencies, one of which is twice degenerate. These two breathing frequencies correspond to two distinguished modes.
In the non-degenerate mode the volume $V(t)$ of the electron gas oscillates, whereas the aspect ratios $\sigma_{i}/\sigma_{j}$ remain constant. This mode preserves the spherical symmetry of the equilibrium state and can be excited by perturbing the three $\sigma_i$ in the same way.
In contrast, for the twice-degenerate mode the volume stays constant whereas the ratios $\sigma_{i}/\sigma_{j}$ oscillate at the corresponding frequency.

These modes are shown in Fig. \ref{fig:linearmodes}. In the figure, we show numerical simulations of the full system obtained by perturbing the stationary ground state by a very small amount $\delta_i$: $\sigma_i(t=0)=\sigma_i^{(0)} + \delta_i$. In all cases, the dipole mode is not excited, i.e., $d_i(t=0)=0$. In the top panel, we only excited the
twice-degenerate mode by choosing the perturbations $\delta_i$ such that the volume is invariant: as expected, the volume stays constant during the linear evolution, while the various $\sigma_i$ oscillate.
In the bottom panel, we only excited the non-degenerate mode by taking $\delta_x=\delta_y=\delta_z$: here, the volume oscillates while the ratio $\sigma_{x}(t)/\sigma_{z}(t)$ remain constant.

In the case of isotropic confinement ($k_x=k_y=k_z\equiv k$), analytical expressions for the dipole frequency $\Omega_d$ and for the degenerate ($\Omega_{\sigma}^{(d)}$) and non-degenerate ($\Omega_{\sigma}^{(nd)}$) breathing frequencies can be found in the large $N$ limit. For harmonic confinement ($\zeta=0$), one obtains the following expressions (which are actually exact for all values of $N$):
\begin{equation}
\Omega_d=\sqrt{k},\,\,\,\,\,\,\, \Omega_{\sigma}^{(nd)}=\sqrt{3k}, \,\,\,\,\,\,\, \Omega_{\sigma}^{(d)}=\sqrt{6k},
\end{equation}
whereas for anharmonic confinement ($\zeta>0$) in the large $N$ limit:
\begin{equation}
\Omega_{d}=\left[\left(20\zeta\right)^{3/2}\alpha_{1}N\right]^{1/5}, \,\,\,\, \Omega_{\sigma}^{(nd)}=5\left[\left(20\zeta\right)^{3/2}\alpha_{1}N\right]^{1/5}, \,\,\, \, \Omega_{\sigma}^{(d)}=\frac{34}{5}\left[\left(20\zeta\right)^{3/2}\alpha_{1}N\right]^{1/5}.
\label{eq:omegad}
\end{equation}
Note that the presence of an anharmonic part in the confining potential introduces a dependence on the number of particles $N$ in the frequencies.

\begin{figure}[h!]
\includegraphics[scale=0.5]{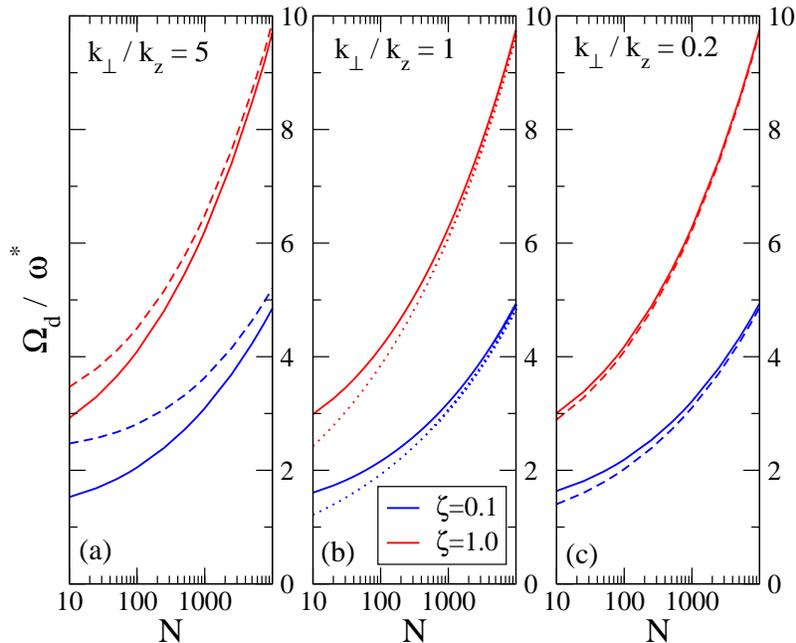}
\caption{\textit{Color online}. Dipole frequencies $\Omega_{d_z}$ (solid lines) and $\Omega_{d_\perp}$ (dashed lines) as a function of the number of electrons $N$ for two values of the anharmonicity parameter, $\zeta=0.1$ (blue) and $\zeta=1$ (red). The left panel (a) corresponds to a cigar-shaped trap with $k_\perp > k_z$; the middle panel (b) to an isotropic trap ($k_\perp > k_z$); and the right panel (c) to a pancake shaped trap ($k_\perp < k_z$). In the isotropic case (b) the longitudinal ($z$) and transverse ($\perp$) dipoles coincide and the dotted lines represent the analytical expressions of Eq. \eqref{eq:omegad}}.
\label{fig:omg-dipole}
\end{figure}

In Fig. \ref{fig:omg-dipole}, we show the dependence of the dipole frequency with
the geometry of the trap (characterized by the parameter $k_\perp/k_z$, where $k_\perp\equiv k_x=k_y$) and the number of electrons. Note that when $k_\perp/k_z \ll 1$ the trap is ``pancake shaped", while in the opposite case $k_\perp/k_z \gg 1$ it is ``cigar shaped"; $k_\perp/k_z = 1$ denotes an isotropic trap.

For an isotropic confinement (Fig. \ref{fig:omg-dipole}, middle panel) the analytical expressions (dotted lines) match closely the numerical results for $N \gg 1$. The anharmonicity introduces a dependence of the dipole frequency with the number of electrons, with higher frequencies corresponding to larger $N$. The same trend is observed for a cigar-shaped trap (left panel) and a pancake-shaped trap (right panel). In these anisotropic traps, the longitudinal (parallel to $z$) and transverse ($\perp$) dipole frequencies of course do not coincide, but both still grow with $N$.

\section{Nonlinear regime and harmonic generation}\label{sec:nonlinear}
In the previous sections, we characterized the linear response of the electron dynamics by studying the eigenvalues of the linearized system of equations. Physically, the linear response corresponds to a weak excitation of the system and results in one or a few lines in the frequency spectrum. In order to trigger high harmonic generation (HHG), it is often necessary to probe the nonlinear response regime, typically by increasing the excitation.

\subsection{HHG and Poincar\'e sections}
In a first set of simulations in the nonlinear regime, we show that HHG is accompanied by some typical signatures of deterministic chaos in the dynamics. Here, we use Poincar\'e sections as evidence of chaotic behavior. We also point out that this type of study is feasible because our reduced mathematical model is a system of ordinary differential equations, which can be analyzed with the usual methods of classical Hamiltonian mechanics.

The method of Poincar\'{e} sections consists in choosing a two-dimensional cross-section (i.e., a plane) of the entire phase space (which, in our case, is six-dimensional) and recording the position on such plane each time that the representative point of the system crosses it.
If the system is chaotic, then there is no correlations between the various points
on the Poincar\'e section, and some finite 2D regions of the plane will be covered uniformly. In contrast, if the system is regular, i.e. periodic, the representative point of the system will pass through the same point after some time, and the Poincar\'e section will consist of isolated points or 1D lines on the plane.

\begin{figure}[h!]
        \centering \includegraphics[scale=0.4,angle=-90]{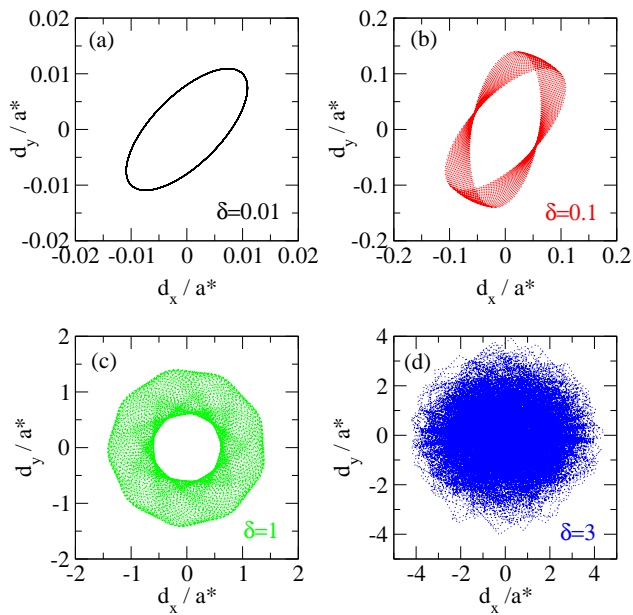}
        \caption{\textit{Color online}. Poincar\'e sections in the plane $(d_x, d_y)$ for different values of the initial excitation: $\delta=0.01$ (a), 0.1 (b), 1 (c), and 3 (d). The simulations were performed for an anisotropic trap with $k_{\perp}=5$, $k_{z}=1$, $N=50$, and $\zeta=0.01$.}
        \label{fig:poincare-delta}
\end{figure}

In our case, we choose the $(d_x, d_y)$ plane for the Poincar\'{e} section. Such plane divides the electron trap in two identical regions. In the forthcoming simulations we take $N=50$ electrons and an anisotropic trap characterized by $k_{\perp}=5$ and $k_{z}=1$.
We perturb the stationary ground state by suddenly changing the position and velocity of the dipole variable, i.e., by setting the following initial conditions at $t=0$: $d_i=\delta$, $\dot{d}_{x}=-\dot{d}_{y}=\delta$, and $\dot{d}_{z}=0$, with the perturbation amplitude $\delta$ varying between 0.01 and 3,
and $\sigma_i(t=0) = \sigma_i^{(0)}$.
The corresponding Poincar\'e sections are shown in Fig. \ref{fig:poincare-delta}.
For $\delta=0.01$ the system is clearly regular, as the Poincar\'e section is basically an ellipse (this is due to the choice of the initial condition).
By increasing $\delta$, the central phase-space region starts filling up, first partially and in a regular way (Fig. \ref{fig:poincare-delta}b-c) and then completely for $\delta=3$ (Fig. \ref{fig:poincare-delta}d). The homogeneous coverage of a finite phase-space area is a signature of chaotic behavior.

\begin{figure}[h!]
\centering \includegraphics[scale=0.4]{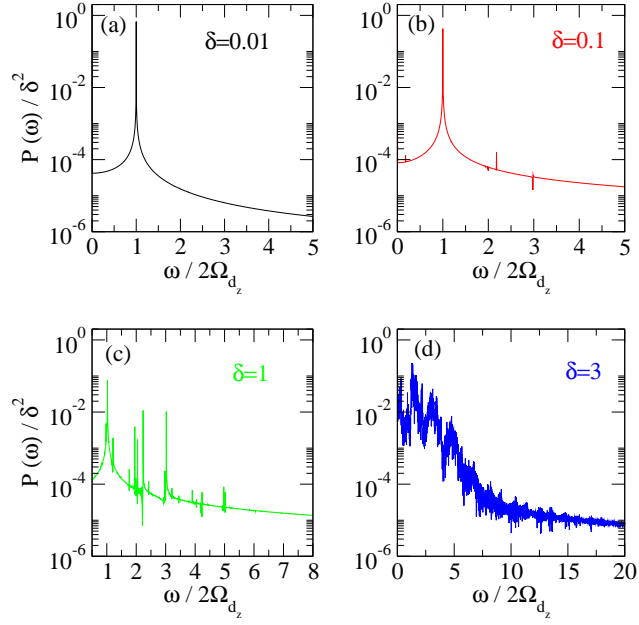}
        \caption{\textit{Color online}. Power spectrum of the dipole $P(\omega)$ normalized to the square of the initial excitation $\delta$, for the same cases as in Fig. \ref{fig:poincare-delta}.}
        \label{fig:spectrum-delta}
\end{figure}

It is interesting to check how the onset of chaos with increasing perturbation correlates with the total power radiated by the electron gas. At large distances, the electron gas can be viewed as an electric dipole of charge $-Ne$ and displacement $d_z(t)$ oscillating along the $z$ axis.
In this case we can apply the Larmor formula \cite{jackson} for the total radiated power: $P(t) = e^2/(6\pi \epsilon_0 c^{3}) |\ddot{d_z}(t)|^2$.
The dipole power spectrum $P(\omega)$ is shown in Fig. \ref{fig:spectrum-delta} for the same cases as in Fig. \ref{fig:poincare-delta}.
As expected, the spectrum displays a single line at the dipole frequency when the excitation is weak ($\delta=0.01$). Higher order harmonics start appearing at larger values of $\delta$, and are at the origin of the multiperiodic motion observed in the corresponding Poincar\'e sections. Finally, for $\delta=3$ the spectrum is nearly continuous, in agreement with the chaotic dynamics observed in Fig. \ref{fig:poincare-delta}.

\begin{figure}[h!]
\centering\includegraphics[scale=0.4]{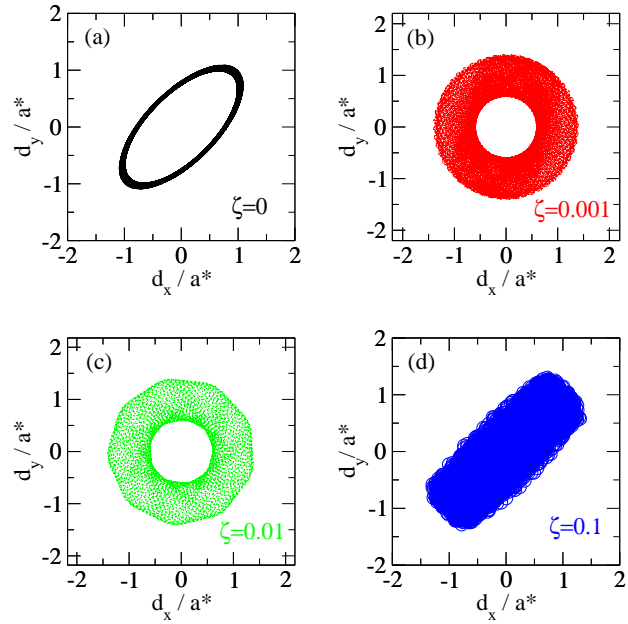}
\caption{\textit{Color online}. Poincar\'e sections in the plane $(d_x, d_y)$ for different values of the anharmonicity parameter $\zeta=0$ (a), 0.001 (b), 0.01 (c), and 0.1 (d). The simulations were performed for an anisotropic trap with $k_{\perp}=5$, $k_{z}=1$, $N=50$, and initial excitation $\delta=1$.}
\label{fig:poincare-zeta}
\end{figure}

\begin{figure}[h!]
\centering \includegraphics[scale=0.4]{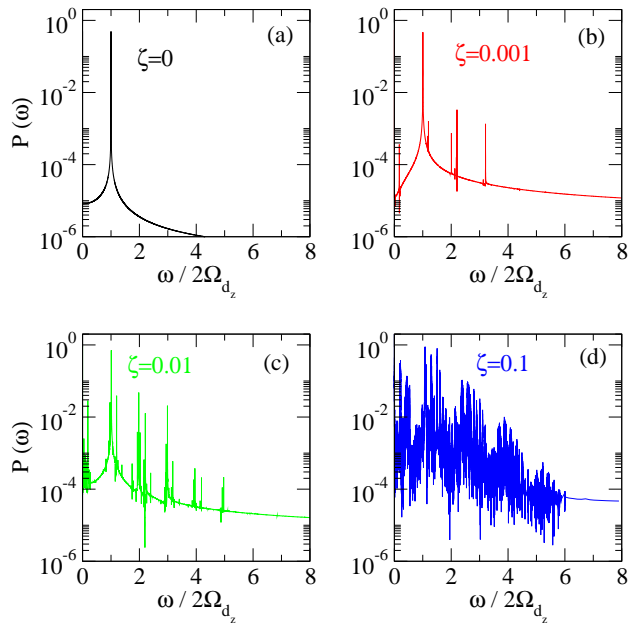}
\caption{\textit{Color online}. Power spectrum of the dipole $P(\omega)$ for different values of the anharmonicity parameter $\zeta$, for the same case as in Fig. \ref{fig:poincare-zeta}.}
\label{fig:spectrum-zeta}
\end{figure}

The same transition to chaos accompanied by HHG was observed for a case where we keep the excitation constant ($\delta=1$) and increase the anharmonicity parameter from $\zeta=0$ to $\zeta=0.1$ (Figs. \ref{fig:poincare-zeta}-\ref{fig:spectrum-zeta}).

To sum up, the situation can be described as follows.
In order to observe some chaotic dynamics, the presence of an anharmonic term in the confinement is necessary -- a purely harmonic oscillator is always integrable.
A finite value of $\zeta$ introduces some coupling between the dipole and the breathing motions, which enlarges the available phase space and allows chaotic behavior.
This chaotic behavior is displayed only when the system explores the nonparabolic regions of the confining trap. This can be achieved either by increasing the initial excitation (Figs. \ref{fig:poincare-delta}-\ref{fig:spectrum-delta}) or by increasing the anharmonicity of the trap (Figs. \ref{fig:poincare-zeta}-\ref{fig:spectrum-zeta}).
Finally, we note that a certain degree of anisotropy ($k_\perp=5k_z$ in our case) was also required to observed such irregular motion.

\subsection{HHG and resonant excitation}
So far, we used a simple excitation for our nonlinear system, namely an initial velocity imparted on the dipole variables $\dot d_i(t=0)$.
In reality, the electron dynamics is usually triggered by electromagnetic (laser) pulses.
In order to simulate this scenario, we assume that the confined electron gas is excited via an oscillating electric field directed along the $z$ axis, $\mathbf{E} = E_{z}(t)\mathbf{e}_z$.
It can be shown that the effect of the field can be included simply by adding a term $-eE_{z} d_z$ to the lagrangian $L$. We consider three cases here: (i) an excitation at a nonresonant frequency, (ii) an excitation at a resonant frequency, and (iii) an excitation with chirp (autoresonance). Note that the dipole linear resonant frequency is in the Tera-Hertz domain.

\begin{figure}[h!]
	\centering \includegraphics[scale=0.5]{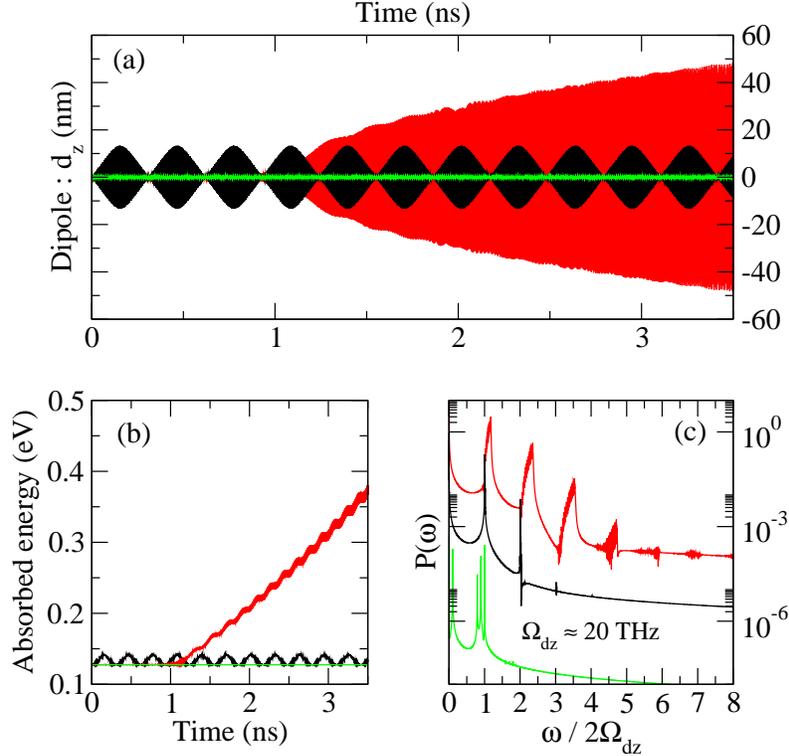}
\caption{\textit{Color online}. Laser excitation of the dipole response, for a trap with parameters: $k_{\perp}=5$, $k_{z}=1$, $N=50$, $\zeta=0.01$. The amplitude of the excitation is $E_0=1.05\times 10^{4}\, \rm V/m$. We show the dipole amplitude $d_z(t)$ (a), the absorbed energy (b), and the power spectrum $P(\omega)$ (c), for three cases: nonresonant excitation at constant frequency (green curves), resonant excitation at constant frequency (black), and chirped excitation (red).}
        \label{fig:laser}
\end{figure}

The results are shown in Fig. \ref{fig:laser}. For the first two cases, the excitation has the form $E_{z}(t) = E_{0} \cos (\omega_{0} t)$, where $E_0$ is the electric field amplitude of the electromagnetic wave. In all cases shown here, we took the same amplitude $E_0= 0.01 \,\rm au$, corresponding to $E_0 =  1.05 \times 10^{4} \,\rm V/m$ in SI units, which can be easily reached experimentally.
In the first case (green curves on the figure), $\omega_0=0.8 \Omega_{d_z}$ differs from the linear response frequency $\Omega_{d_z}$. Being out of resonance, the system stays close to the linear regime: the oscillation amplitude remains small (Fig. \ref{fig:laser}a) and only a small amount of energy is absorbed by the electron gas (Fig. \ref{fig:laser}b)
The power spectrum (Fig. \ref{fig:laser}c) displays four (small) peaks, corresponding to the laser frequency, the linear response frequency, and the harmonics $\Omega_{d_z}\pm \omega_0$.

For a resonant excitation $\omega_0=\Omega_{d_z}$ (black curves), the oscillation amplitude and the absorbed energy initially increase, but then decrease again after some time. This is because the effective force acting on the dipole is not harmonic and the resonant frequency actually depends on the amplitude of the oscillations.
When the amplitude grows and the system reaches the nonlinear regime, the fixed external frequency no longer matches the instantaneous resonant frequency (which differs from $\Omega_{d_z}$ in the nonlinear regime). The resulting power spectrum
displays two lines corresponding to the linear frequency and the first harmonic.

Finally, we use an oscillating field with a chirped frequency:
$E_{z}(t) = E_{0} \cos \left(\omega_{0} t + \frac{\alpha}{2} t^{2}\right)$, where $\alpha$ is the rate of variation of the laser frequency.
This type of forcing is known as autoresonance \cite{auto} and was applied in the past to many physical systems \cite{Fajans99,Murch11}.
Autoresonance occurs when a classical nonlinear oscillator is externally excited by an oscillating field with slowly varying frequency.
For $|\alpha| \ll \omega_0^2$ (adiabatic process) and
$E_0$ above a certain threshold, the instantaneous oscillator
frequency becomes ``locked" to the instantaneous excitation
frequency, so that the resonance condition is always satisfied. In
that case, the amplitude of the oscillations grows indefinitely
and without saturation, until of course some other effect becomes dominant.
Usually the threshold behaves as $E_0^{\rm th} \sim |\alpha|^{3/4}$, so that the
amplitude can be arbitrarily small provided that the
external frequency varies slowly enough \cite{auto}.

In an earlier study, using the autoresonant technique in conjunction with a phase-space model of the electron dynamics, we showed that it is possible to efficiently extract the electrons from a Gaussian-shaped quantum well \cite{ap_auto4}.
More recently, autoresonant excitation was used to trigger HHG in metallic nanoparticles \cite{hurst_harmo}. We now show that it can be very effective also for the systems considered here, namely quantum dots and quantum wells.

In order for autoresonance to work, the excitation frequency, which varies linearly in time as $\omega(t)=\omega_{0} + \alpha t$, must cross at some point the resonant dipole frequency $\Omega_{d_z}= 20.07\, \rm THz$. Therefore, for this simulation we chose $\omega_0 = 18.45\, \rm THz$ and a chirp rate $\alpha = 3.03\, \rm THz/ns$. The resonant frequency is crossed around $t \approx 0.55\, \rm ns$, after which the autoresonant process starts being effective, as can be seen in Fig. \ref{fig:laser}.
It is clear from Fig. \ref{fig:laser} that the autoresonant excitation allows one to increase phenomenally the amplitude of the dipole oscillations and consequently the absorbed energy, which is roughly three times as large compared to the non chirped case. We stress that the excitation amplitude is the same for all cases.
The power spectrum (Fig. \ref{fig:laser}c) displays several peaks for higher order harmonics (up to the third harmonic), with the first harmonic being roughly a factor of ten smaller that the linear mode.
We also note that these spectral lines are unusually broad. This is probably due to the chirped excitation, which sweeps several frequencies around each harmonic.

The important point is that, using a rather weak excitation ($E_0 = 1.05\times 10^4\, \rm V/m$  in the present case, which could be made even weaker by reducing the chirp rate $\alpha$), one can induce significant energy absorption by the electron gas, accompanied by HHG at remarkably high levels.

\section{Conclusions}
HHG is a highly topical research area with many potential applications \cite{nappa}; most notably it is a prerequisite for the generation of attosecond laser pulses \cite{attosecond}.
In this work, our aim was to explore the possibility of HHG using nanometric system containing many electrons, such as semiconductor quantum dots and wells.
With this purpose in mind, we constructed an effective model in the form of a dynamical system made of six coupled differential equations for the center of mass and the size of the electron gas. This effective model results from the application of a variational method to the equations of quantum hydrodynamics.

The model was later applied to the dynamics of an electron gas in a nonparabolic and anisotropic well. Two main results were obtained. First, we showed that harmonic generation is accompanied by dynamical chaos in the equations of motion. The onset of chaos was quantified by the appearance of ergodic regions in some Poincar\'e sections.

Second, we demonstrated that HHG can be efficiently achieved by exciting the system with a chirped laser pulse. This process, known as classical autoresonance, is capable of bringing the electrons into a strongly nonlinear regime, leading to the generation of high harmonics. Crucially, the autoresonance technique works well for relatively modest driving fields and does not require any fine tuning of the laser pulse.

The present results complete and extend to three spatial dimensions our earlier findings that HHG can be triggered with similar techniques in systems of metallic nanoparticles \cite{hurst_harmo}.

\vskip 5mm
{\bf Acknowledgements}\\
We thank the Agence Nationale de la Recherche, project Labex ``Nanostructures in Interaction with their Environment", for financial support.
FH acknowledges support from the Brazilian research fund CNPq (Conselho Nacional de Desenvolvimento Cient\'ifico e Tecnol\'ogico-Brasil).

%%%%%%%%%%%%%%%%%%%%%%%%%%%%%%%%%%%%%%%%%%%%%%%%%


\begin{thebibliography}{9}

\bibitem{Zoller}P. Zoller et al., Eur. J. Phys. {\bf 36}, 203
(2005).

\bibitem{Kohn} W. Kohn, Phys. Rev. {\bf 123}, 1242 (1961).
\bibitem{Dobson} J. F. Dobson, Phys. Rev. Lett. {\bf 73}, 2244
(1994).

\bibitem{sako} T. Sako, P.-A. Hervieux, and G. H. F. Dierksen, Phys. Rev. B {\bf 74},
045329 (2006).

\bibitem{schroter} S. Schr\"oter, P.-A. Hervieux, G.
Manfredi, J. Eiglsperger, and J. Madro$\tilde{\rm n}$ero, Phys. Rev. B {\bf 87}, 155413 (2013).

\bibitem{Muller} T. M{\"u}ller, W. Parz, G. Strasser, and K.
Unterrainer, Phys. Rev. B {\bf 70}, 155324 (2004)

\bibitem{Pereira} M. F. Pereira and H. Wenzel, Phys. Rev. B {\bf 70}, 205331 (2004).

\bibitem{Nikonov} D. E. Nikonov, A. Imamoglu, L.
V. Butov, and H. Schmidt, Phys. Rev. Lett. {\bf 79}, 4633 (1997).
\bibitem{Ullrich} H. O. Wijewardane and C. A. Ullrich, Appl. Phys. Lett. {\bf
84}, 3984 (2004).

\bibitem{ap_auto4} G. Manfredi and P.-A. Hervieux, Appl. Phys. Lett. \textbf{91}, 061108 (2007).

\bibitem{fernando_book} F. Haas, \emph{Quantum plasmas an hydrodynamic approach} (Springer, Berlin, 2011).

\bibitem{qhy} G. Manfredi and F. Haas, Phys. Rev. B \textbf{64}, 075316 (2001).

\bibitem{qhy_mol} M. Brewczyk, K. Rzazewski, and C. W. Clark, Phys. Rev. Lett. \textbf{78}, 191 (1997).

\bibitem{qhy_clust1} A. Banerjee, M. K. Harbola, J. Chem. Phys. \textbf{113}, 5614 (2000).

\bibitem{qhy_clust2} A. Domps, P.-G. Reinhard, and E. Suraud, Phys. Rev. Lett. \textbf{81}, 5524 (1998).

\bibitem{manfredi2012} G. Manfredi, P. A. Hervieux, and F. Haas, New J. Phys. \textbf{64}, 075012 (2012).

\bibitem{ciraci} C. Cirac\`{\i},  J. B. Pendry, and D. R. Smith, Chem. Phys. Chem. {\bf 14}, 1109 (2013).

\bibitem{qhy_thin_met} N. Crouseilles, P.-A. Hervieux, and G. Manfredi, Phys. Rev. B \textbf{78}, 155412 (2008).

\bibitem{qhy_semi_cond} F. Haas, G. Manfredi, P. K. Shukla, and P.-A. Hervieux, Phys. Rev. B \textbf{80}, 073301 (2009).

\bibitem{hurst_harmo}J. Hurst, F. Haas, G. Manfredi, and P.-A. Hervieux, Phys. Rev. B \textbf{89}, 161111(R) (2014).

\bibitem{Bonitz2015} D. Michta, F. Graziani, and M. Bonitz,
Contrib. Plasma Phys. \textbf{55}, 437 (2015).

\bibitem{corr_pot} A. D. Becke, Phys. Rev. A \textbf{38}, 3098 (1988).

\bibitem{jackson} J. D. Jackson, \emph{Classical electrodynamics} (Wiley, New York, 1998).

\bibitem{auto} J. Fajans and L. Friedland, Am. J. Phys. \textbf{69}, 1096 (2001).

\bibitem{Fajans99} J. Fajans, E. Gilson, and L. Friedland, Phys. Rev. Lett. \textbf{82}, 4444 (1999).

\bibitem{Murch11} K. W. Murch et al., Nature Phys. \textbf{7}, 105 (2011).

\bibitem{nappa} J. Nappa, G. Revillod, I. Russier-Antoine, E. Benichou, C. Jonin, and P. F. Brevet, Phys. Rev. B {\bf 71}, 165407 (2005).

\bibitem{attosecond} Ferenc Krausz and Misha Ivanov, Rev. Mod. Phys. {\bf 81}, 163 (2009).

\end{thebibliography}
\end{document}